\documentclass[3p,times,twocolumn]{elsarticle}
\usepackage{empheq}
\usepackage[skins,theorems,many]{tcolorbox}
\usepackage{bm}
\usepackage{algorithm}
\usepackage[noend]{algpseudocode}
\usepackage{calrsfs}
\usepackage{caption}
\usepackage{color}
\usepackage[dvipsnames]{xcolor}
\usepackage{wrapfig}
\usepackage{booktabs}
\usepackage{lscape}
\usepackage{float}
\usepackage{flafter}
\usepackage{epstopdf}
\usepackage{subcaption}
\usepackage{amsmath}
\usepackage{amssymb}
\usepackage{siunitx}
\usepackage{mathrsfs}
\usepackage[makeroom]{cancel}
\usepackage{tikz}
\usepackage{soul} 
\usepackage{xr}
\usepackage{dblfloatfix}
\journal{Extreme Mechanics Letters}

\begin{document}

\begin{frontmatter}

\title{Meso-scale modeling of cohesive cavitation in polymer networks}

\author[CU]{Saleh Assadi\fnref{equal}}
\author[CU]{Zachary T. White\fnref{equal}}
\author[3M]{Fay Salmon}
\author[3M]{Nitin Hansoge}
\author[3M]{Zhuonan Liu}
\author[3M]{Victor Crespo-Cuevas}
\author[3M]{Jason Clapper}
\author[Cornell,pasteur]{Nikolaos Bouklas}
\author[CU]{Franck J. Vernerey\corref{cor1}}
\ead{franck.vernerey@colorado.edu}

\fntext[equal]{These authors contributed equally to this work.}
\cortext[cor1]{Corresponding author}

\affiliation[CU]{organization={Department of Mechanical Engineering, University of Colorado},
            city={Boulder}, state={CO}, country={USA}}
\affiliation[3M]{organization={Corporate Research Lab, 3M Company},
            city={St. Paul}, state={MN}, country={USA}}

\affiliation[Cornell]{
    organization={Sibley School of Mechanical and Aerospace Engineering, Cornell University},
    city={Ithaca}, state={NY}, country={USA}
}

\affiliation[pasteur]{
    organization={Pasteur Labs},
    city={Brooklyn}, state={NY}, country={USA}
}

\begin{abstract}
Cavitation influences the failure of elastomers and soft adhesives, but connecting its microscopic initiation to subsequent cavity growth remains challenging. Here, we develop a discrete mesoscale framework that combines finitely extensible, breakable strands with local van der Waals interactions to examine how cohesion and network heterogeneity transform distributed dilation into localized damage. In the simulations, local cohesive instability produces a network-scale precursor whose subsequent expansion is coupled to strand stretching and rupture. Cohesive strength and intrinsic defect size govern the critical hydrostatic tension, while cohesion also controls the number and spatial distribution of cavities. Under confined loading, decreasing the layer thickness promotes multiple cavities and distributed strand rupture, extending the load-bearing response and increasing the critical work density of rupture. An energetic interpretation based on the thickness dependence of the crack-driving force connects these observations to competition between additional cavity nucleation and propagation of existing damage. The results link local cohesive instability to the collective evolution of cavities and rupture in disordered polymer networks.
\end{abstract}

\begin{keyword}
Cavitation \sep Soft Materials \sep Computational Mechanics \sep Molecular Cohesion \sep Network Architecture
\end{keyword}

\end{frontmatter}

\section{Introduction}

Soft polymer networks can fail differently depending on loading conditions, with deformation accommodated by stretching, internal cavity growth, and fracture. Cavitation is particularly important in bonded rubbers and soft adhesives, where confinement restricts lateral contraction and makes the response to volumetric deformation central to failure. Bonded-rubber and poker-chip experiments have long demonstrated the formation of internal cavities under tension~\cite{Gent1957,Gent1959}, while studies of pressure-sensitive adhesives show how cavities develop alongside fibrillation and debonding~\cite{Lakrout01031999,Creton_2016}. Understanding these processes requires connecting the formation of a small cavity within the network to its subsequent growth and interaction with the surrounding material.

Classical elastic descriptions start with the expansion of a pre-existing void under tension. For an infinitesimal spherical void in an infinite, incompressible neo-Hookean solid, the classical instability occurs at a remote hydrostatic tension of $5G/2$, where $G$ is the shear modulus~\cite{Gent1959}, although this result depends on the constitutive and geometric assumptions and does not define a universal cavitation strength~\cite{Gent1990CavitationIR}. Nonlinear-elastic theories extend this picture to cavitation from vanishingly small defects, finite defect populations, and multiaxial loading~\cite{Ball1982,MullerSpector1995,Horgan1995,LOPEZPAMIES20111464,LOPEZPAMIES20111488}. These approaches establish conditions for elastic cavity growth, but the size and physical origin of the initial defect remain important for interpreting experiments, which distinguish gradual enlargement of visible defects from abrupt cavity formation at initially unresolved sites~\cite{DOLLHOFER20046111}. Surface energy further increases the resistance to expansion of small defects, while the surrounding geometry determines the elastic energy available for growth~\cite{Gent_tompkins_surface_energy,Chiche2005}.

A complementary view treats cavity growth as a fracture process~\cite{Williams1965}, motivated by experiments showing that cavitation thresholds depend on fracture resistance as well as elastic modulus~\cite{Cristiano2010} and by cavity-expansion measurements that reveal contributions from both elasticity and fracture~\cite{KunduCrosby2009,Hutchens2016,RaayaiArdakani2019}. Theoretical descriptions identify transitions between elastic expansion and crack-like growth~\cite{Kang2017,Fu2023}, while phase-transition formulations distinguish strength-controlled nucleation from energy-controlled propagation in confined elastomers~\cite{KUMAR2018523,kumar2018configurational,KUMAR2021104359,KAMAREI2024105683}. Observations of cavities developing into microcracks and macroscopic cracks~\cite{Lefevre2015,Poulain2017}, together with molecular damage surrounding apparently smooth cavities~\cite{Morelle2021,Barney2022Linking}, indicate that cavity shape alone cannot identify the underlying growth mechanism. Internally nucleated microcracks can also close and heal after unloading~\cite{Poulain2018}, showing that the relationship between cavity opening and irreversible damage depends on the material and loading history. In confined layers, the progression toward macroscopic failure also depends on thickness, with thinner specimens supporting multiple cavities over a wider loading interval and thicker specimens tending toward a dominant failure event~\cite{Chiche2005,LeMenn2022}.

Connecting these continuum descriptions to cavity initiation requires considering the network scale, where molecular cohesion, structural disorder, and strand deformation interact. Elastomers are nearly incompressible~\cite{Pritchard2013,DOGRU2018375}, although their bulk response can depend on network structure~\cite{BARNEY2022101616}. Their strand elasticity is primarily entropic~\cite{James1047,Edwards_1967,Flory1979MolecularTO}, while cohesive interactions between molecular groups provide much of the resistance to local dilation~\cite{TABOR19942759}. Molecular simulations resolve nanoscale voids in crosslinked networks~\cite{ZEE201555} and distinguish the pressure required to nucleate a cavity from that needed to sustain its growth~\cite{Riggleman2020}, highlighting processes that an incompressible elastic description does not explicitly represent. The crosslinked van der Waals framework separates entropic network elasticity from cohesive resistance to dilation~\cite{cohesive_instability_sam}, providing a basis for examining how a local loss of cohesive stability interacts with network rearrangement and strand failure.

Here, we develop a discrete mesoscale model that combines finitely extensible, breakable strands with local-density-dependent van der Waals interactions to connect cavity formation within a disordered network to subsequent growth and rupture. Starting from a dense network without a prescribed continuum-scale cavity, we examine how cohesive strength and intrinsic defects determine the onset and spatial distribution of cavitation, then follow the resulting damage under confined loading. By varying the layer thickness at fixed network parameters, we investigate how the competition between additional cavity formation and propagation of existing damage shapes the cavity population, stress--stretch response, and work required for rupture.

\section{Discrete modeling of a crosslinked network}
\label{sec:discrete_model}

\begin{figure*}[t!]
    \centering
    \includegraphics[width=1\linewidth, trim={1.0cm 0 0 2.0}]{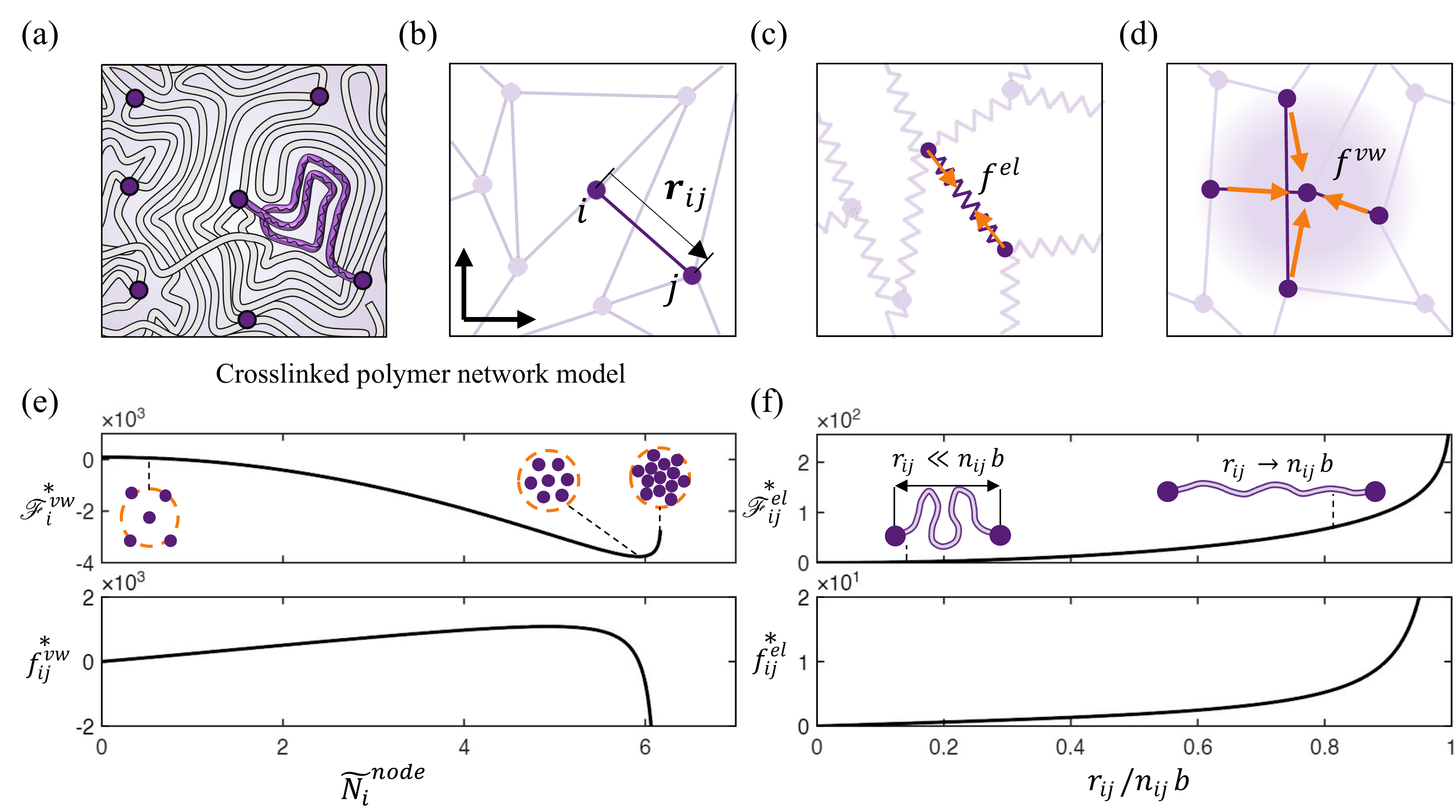}
    \caption{ Discrete representation of a crosslinked polymer network. (a) Molecular-scale depiction of polymer strands connected by crosslinking junctions. (b) Coarse-grained representation of junctions $i$ and $j$, connected by a strand with end-to-end vector $\bm r_{ij}$. (c,d) Entropic strand and local-density cohesive forces. (e) Normalized local van der Waals free energy ($\mathcal{F}_{i}^{\star vw}=\mathcal{F}_{i}^{vw}/k_BT$) and corresponding normalized cohesive force ($f^{\star vw}_{ij} = f^{vw}_{ij}b/k_BT$) as functions of the weighted local node count. (f) Entropic free energy and tensile force of a polymer strand as functions of its normalized extension. Starred energies and forces are normalized by $k_BT$ and $k_BT/b$, respectively.}
    \label{fig:network_depiction}
\end{figure*}

The discrete model translates the crosslinked van der Waals fluid description of Lamont et al.~\cite{cohesive_instability_sam} to a structurally disordered polymer network. The molecular-scale schematic shown in Fig.~\ref{fig:network_depiction}(a) consists of polymer strands constrained by crosslinking junctions within a densely packed polymer phase. At the mesoscale, this architecture is coarse-grained into the network shown in Fig.~\ref{fig:network_depiction}(b). Each crosslinking junction is represented by a node with position $\bm r_i$, while each polymer strand connecting nodes $i$ and $j$ is represented by the bond vector $\bm r_{ij}=\bm r_j-\bm r_i$, with length $r_{ij}=\lVert\bm r_{ij}\rVert$. This representation retains an explicit network length scale and allows density fluctuations, non-affine deformation, cavity formation, and strand rupture to emerge without prescribing an initial large defect. However, here we do not consider the effects of physical entanglements and dynamic reptation of chains \cite{Assadi_soft_matter,Reptation_deGennes}.

Two distinct energetic contributions govern the network response. Connected nodes interact through nonlinear entropic strand forces, as depicted in Fig.~\ref{fig:network_depiction}(c), whereas nodes within a prescribed cutoff neighborhood contribute to the local cohesive force illustrated in Fig.~\ref{fig:network_depiction}(d). The strand contribution controls distortional elasticity, finite extensibility, and rupture, while the cohesive contribution represents the excluded-volume and attractive interactions within the densely packed polymer phase. The total Helmholtz free energy is 
\begin{equation}
\label{eq:total_discrete_energy}
\mathcal F=\sum_{\langle i,j\rangle\in\mathcal B} \mathcal F_{ij}^{el} + \sum_i\mathcal F_i^{vw},
\end{equation}
where $\mathcal B$ denotes the set of polymer strands. Because both contributions depend on the same nodal positions, local dilation, strand stretching, and non-affine network rearrangement remain mechanically coupled.

\subsection{Nonlinear elasticity of polymer strands}
\label{subsec:strand_elasticity}

A strand connecting nodes $i$ and $j$ contains $n_{ij}$ Kuhn segments of length $b$ and therefore has contour length $n_{ij}b$. Its normalized extension is defined as $\lambda_{ij}=r_{ij}/(n_{ij}b)$, where $0\leq\lambda_{ij}<1$. Finite chain extensibility is represented using the Pad\'e approximation to the inverse Langevin function~\cite{Cohen1991}. Taking the coiled state as the reference, the entropic free energy of the strand is
\begin{equation}
\label{eq:chain_energy}
\mathcal F_{ij}^{el}=n_{ij}k_BT\left[\frac{\lambda_{ij}^{\,2}}{2}-\ln\!\left(1-\lambda_{ij}^{\,2}\right)\right].
\end{equation}
The force exerted on node $i$ by strand $i$--$j$, represented schematically in Fig.~\ref{fig:network_depiction}(c), follows directly from the gradient of Eq.~\eqref{eq:chain_energy}:
\begin{equation}
\label{eq:chain_force}
\bm f_{ij}^{el}=-\nabla_{\bm r_i}\mathcal F_{ij}^{el} = \frac{k_BT}{b} \frac{\lambda_{ij} \left(3-\lambda_{ij}^{\,2}\right)} {1-\lambda_{ij}^{\,2}} \frac{\bm r_{ij}}{r_{ij}},
\end{equation}
with $\bm f_{ji}^{el}=-\bm f_{ij}^{el}$. Figure~\ref{fig:network_depiction}(f) plots the resulting normalized strand energy $\mathcal F_{ij}^{el}/k_BT$ and tensile force $f_{ij}^{el}b/k_BT$ versus normalized extension $\lambda_{ij}$. In the Gaussian limit, $\lambda_{ij}\ll1$, the energy is approximately quadratic and the force reduces to $\bm f_{ij}^{el}\simeq (3k_BT/n_{ij}b^2)\bm r_{ij}$. As $\lambda_{ij}\rightarrow1$, both the free energy and tensile force increase rapidly, capturing the finite-extensibility stiffening that occurs as a strand approaches its contour length.

To represent irreversible strand failure while avoiding the numerical singularity at full extension, a strand is permanently removed when $\lambda_{ij}\geq\lambda_{\mathrm{rup}}=0.95$. The network is then allowed to redistribute the released force. This criterion enables the model to describe entropic elasticity before failure and rupture-mediated cavity growth at large local deformation.

\subsection{Van der Waals cohesive interactions}
\label{subsec:vdw_interactions}

The strand energy describes the distortional elasticity of the network but does not account for the large resistance of the densely packed polymer phase to changes in volume. We therefore supplement the strand elasticity with a local representation of the van der Waals free energy introduced by Johnston ~\cite{VanDerWaals_Johnston_2014}. The volume-occupying particles are identified with Kuhn segments, each occupying a characteristic volume $\nu\sim b^3$. In the present two-dimensional simulations, $\nu$ is interpreted as the corresponding effective segment area.

For $N_p$ Kuhn segments occupying a volume $V$, the van der Waals free energy is~\cite{VanDerWaals_Johnston_2014}
\begin{equation}
\label{eq:vdw_particle_energy}
\begin{aligned}
\mathcal F^{vw}(V,N_p)
={}&-N_pk_BT
\left[
\ln\!\left(\frac{n_q(V-N_p\nu)}{N_p}\right)+1
\right]\\
&-\epsilon_a\frac{N_p^2\nu}{V}.
\end{aligned}
\end{equation}
Here, $\epsilon_a$ sets the attractive energy scale between neighboring Kuhn segments, and $n_q$ is the quantum concentration, with units of inverse volume, which makes the logarithmic argument dimensionless. At fixed temperature, $n_q$ is taken to be a spatially uniform constant. The logarithmic term combines particle entropy with finite-volume exclusion, whereas the second term represents cohesive attraction. Equation~\eqref{eq:vdw_particle_energy} is the fundamental form of the van der Waals energy used in the continuum theory of Lamont et al.~\cite{cohesive_instability_sam}. This free energy function has a non-convex behavior where, during volumetric expansion, the energy rises until the branch loses convexity and the system becomes unstable. These instabilities will lead to localization of cohesive forces and formation of cavity precursors.

For reference, the corresponding continuum free-energy density based on equation~\eqref{eq:vdw_particle_energy} is derived by Lamont et al.~\cite{cohesive_instability_sam} as
\begin{equation}
\label{eq:continuum_vdw_energy}
\psi^{vw}(J)=-\frac{G}{\chi}\left[\ln(J-f_0)+\widetilde{\epsilon}_a\frac{f_0}{J}\right],\qquad\widetilde{\epsilon}_a=\frac{\epsilon_a}{k_BT},
\end{equation}
where $J=\det\bm F=V/V_0$ is the volume ratio, $f_0=N_p\nu/V_0$ is the reference volume fraction occupied by Kuhn segments, and $\chi=V_0c_0/N_p$ is the ratio between the number of network strands and the number of volume-occupying Kuhn segments. Here, $c_0$ is the reference strand density and $G=c_0k_BT$ is the network
shear modulus.

The reference volume fraction is not an independent parameter, since the undeformed configuration must be stress-free. The total continuum free energy $\psi=\psi^{el}+\psi^{vw}$ therefore satisfies
\begin{equation}
\label{eq:reference_equilibrium}
\left.\frac{\partial\psi}{\partial\bm F}\right|_{\bm F=\bm I}=\bm 0,
\end{equation}
which, for the isotropic reference state, reduces to
\begin{equation}
\left.\left(\frac{\partial\psi^{el}}{\partial J}+\frac{\partial\psi^{vw}}{\partial J}\right)\right|_{J=1}=0.
\end{equation}
In the densely packed, long-chain limit, this equilibrium condition requires $\epsilon_a>4k_BT$ and selects the dense branch $1/2<f_0<1$~\cite{cohesive_instability_sam}.

We implement Eq.~\eqref{eq:vdw_particle_energy} using a local-density multibody potential of the form introduced by Sanyal and Shell~\cite{Sanyal16,Sanyal18}. The original discrete implementation of the crosslinked van der Waals model evaluated local density from Voronoi volumes~\cite{cohesive_instability_sam}. Here, a smooth local-density estimate is used during large non-affine motions, cavity opening, and changes in neighboring nodes. The cohesive interaction depends on all nodes within the local neighborhood, including nodes that are not connected by polymer strands [Fig.~\ref{fig:network_depiction}(d)].

For each node $i$, let $\mathcal N_i$ denote the set of nodes within a cutoff radius $r_c$. The weighted local node count is
\begin{equation}
\label{eq:weighted_node_count}
\widetilde N_i^{\,\mathrm{node}}=\sum_{j\in\mathcal N_i}\widehat{\phi}(r_{ij}),\qquad\widehat{\phi}(r)=\frac{\phi(r)}{\langle\phi\rangle_c},
\end{equation}
where $\langle\phi\rangle_c$ is the spatial average of $\phi$ over the cutoff neighborhood. The compact weighting function is
\begin{equation}
\label{eq:density_kernel}
\phi(r)=\begin{cases} a_0+a_2r^2+a_4r^4+a_6r^6, & r<r_c,\\[2pt] 0, & r\geq r_c,
\end{cases}
\end{equation}
with coefficients chosen such that the kernel and its required derivatives smoothly approach zero at $r_c$. The kernel is shown in Fig. S1 of the Supporting Information. For a spatially uniform node distribution, this normalization recovers the corresponding local node count, with division by $V_c$ giving the node density.

The conversion from node count to Kuhn-segment count uses the total number of Kuhn segments in the initial network. Each crosslinking node is assigned the same segment count
\begin{equation}
\label{eq:kuhn_per_node}
\kappa
=
\frac{N_K^{\mathrm{tot}}}{N_{\mathrm{node}}}
=
\frac{1}{N_{\mathrm{node}}}
\sum_{\langle i,j\rangle\in\mathcal B_0} n_{ij},
\end{equation}
where $N_{\mathrm{node}}$ is the total number of crosslinking nodes and $\mathcal B_0$ is the initial strand set, with each strand counted once. This global assignment is independent of local connectivity and remains fixed after strand rupture, which removes an elastic connection without removing the represented polymer material. The weighted number of Kuhn segments in the neighborhood centered on node $i$ is therefore
\begin{equation}
\label{eq:local_kuhn_count}
N_i^K
=
\kappa\sum_{j\in\mathcal N_i}\widehat{\phi}(r_{ij})
=
\kappa\widetilde N_i^{\,\mathrm{node}}.
\end{equation}

Replacing the particle number and volume in Eq.~\eqref{eq:vdw_particle_energy} by the local quantities $N_i^{K}=\kappa\widetilde N_i^{\,\mathrm{node}}$ and $V_c$ gives
\begin{equation}
\label{eq:local_vdw_energy}
\begin{aligned}
\mathcal F_i^{vw}={}&-\kappa\widetilde N_i^{\,\mathrm{node}}k_BT\left[\ln\!\left(\frac{(V_c-\kappa\widetilde N_i^{\,\mathrm{node}}\nu)}{\kappa\widetilde N_i^{\,\mathrm{node}}}\right)+1\right]\\
&-\epsilon_a\frac{\left(\kappa\widetilde N_i^{\,\mathrm{node}}\right)^2\nu}{V_c}.
\end{aligned}
\end{equation}
where $V_c$ is the volume associated with the cutoff neighborhood and $N_i^{K}\nu<V_c$.

\begin{figure*}[t!]
    \centering
    \includegraphics[width=0.9\linewidth, trim={1.0cm 0 0 0}]{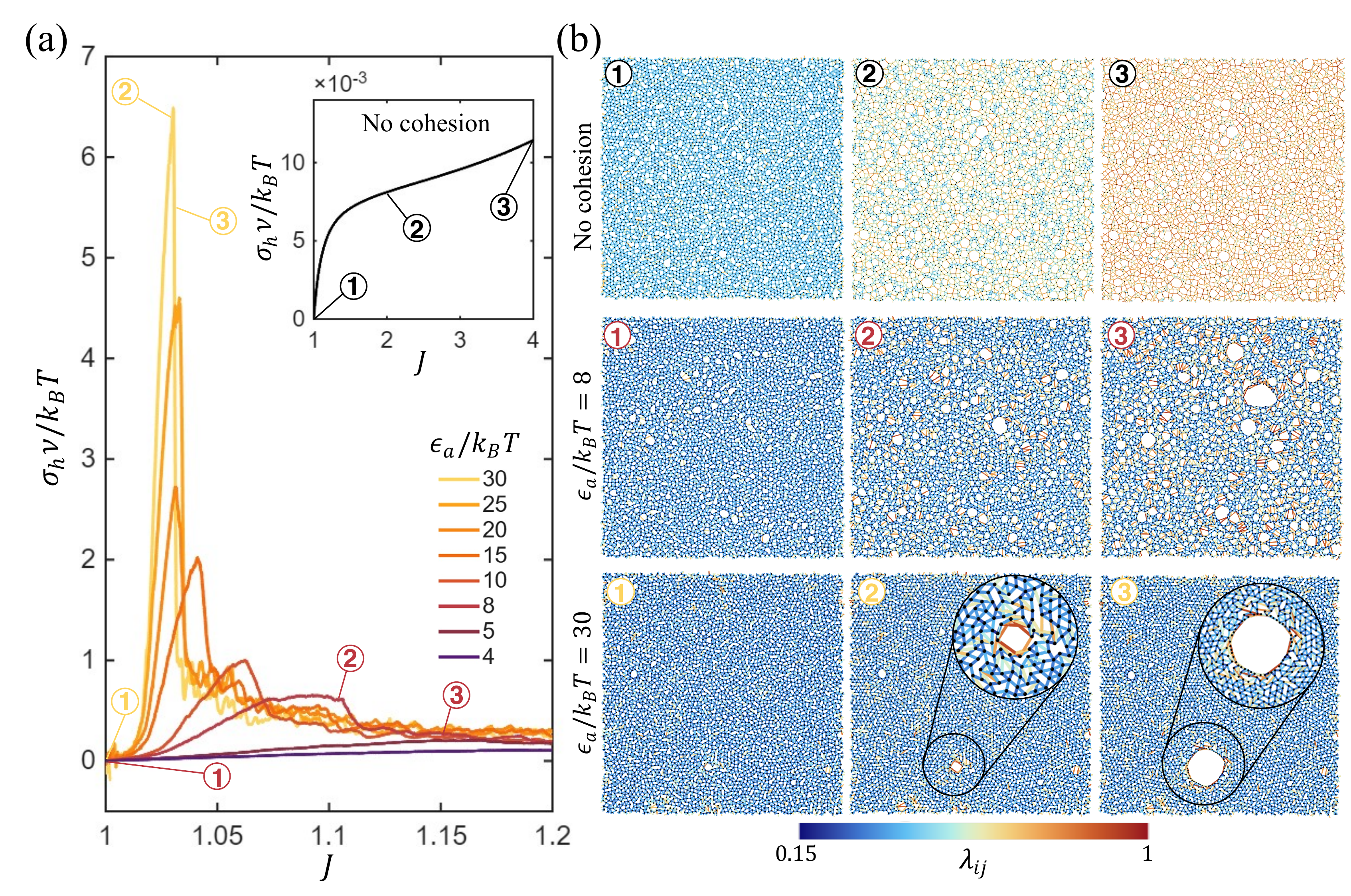}
    \caption{Cohesion transforms swelling to localized cavitation. (a) Hydrostatic stress versus dilation ratio responses for the indicated cohesive strengths and the noncohesive network. (b) Configurations at states 1--3 for no cohesion, $\epsilon_a/k_BT=8$ and $\epsilon_a/k_BT=30$. Strands are colored by $\lambda_{ij}=r_{ij}/(n_{ij}b)$.}
    \label{fig:cohesion_onset}
\end{figure*}

Figure~\ref{fig:network_depiction}(e) illustrates the normalized local cohesive energy $\mathcal F_i^{vw}/k_BT$ versus weighted node count $\widetilde N_i^{\,\mathrm{node}}$, together with the normalized cohesive force $f_{ij}^{vw}b/k_BT$. The energy possesses a minimum that defines the locally equilibrated, densely packed state. Increasing the node count beyond this minimum activates the steep excluded-volume repulsion. Conversely, decreasing $\widetilde N_i^{\,\mathrm{node}}$ moves the local environment onto the dilational branch, where the energy eventually loses convexity. At this point the cohesive force begins to drop, providing the local cohesive instability from which a cavity precursor can emerge.

The local energy $\mathcal F_i^{vw}$ is associated with the neighborhood centered on node $i$, while the total cohesive energy is the sum of these node-centered contributions. A change in the separation $r_{ij}$ changes both the contribution of node $j$ to the neighborhood centered on $i$ and the contribution of node $i$ to the neighborhood centered on $j$. Differentiating the total cohesive energy therefore gives
\begin{equation}
\label{eq:local_vdw_pair_force}
\bm f_{ij}^{vw}
=
\kappa
\left(
\frac{\partial\mathcal F_i^{vw}}{\partial N_i^K}
+
\frac{\partial\mathcal F_j^{vw}}{\partial N_j^K}
\right)
\widehat{\phi}'(r_{ij})
\frac{\bm r_{ij}}{r_{ij}}.
\end{equation}
The two derivatives describe the energetic response of the neighborhoods centered on $i$ and $j$ to changes in their weighted Kuhn-segment counts. Their combined contribution ensures $\bm f_{ji}^{vw}=-\bm f_{ij}^{vw}$, and the total cohesive force on node $i$ is $\bm f_i^{vw}=\sum_{j\in\mathcal N_i,\,j\ne i}\bm f_{ij}^{vw}$. Although written as pair contributions, these forces depend on the surrounding local densities and therefore represent a multibody interaction. The numerical energy-minimization scheme is described in the Supporting Information.

We assessed the local cohesive potential against the homogeneous continuum theory of Lamont et al.~\cite{cohesive_instability_sam} under isotropic expansion [Fig. S3(a)]. The discrete networks reproduce the main features of the pre-instability hydrostatic stress--dilation response and the increase in bulk modulus with cohesive energy, although their bulk modulus is systematically lower. This difference may reflect the smoothed local-density representation and the non-affine relaxation available to the discrete network. The critical hydrostatic tension is also lower than the homogeneous prediction, consistent with intrinsic variations in local density and strand arrangement allowing a weak region to become unstable before the corresponding homogeneous state.

\section{Results and discussion}
\label{sec:results}

Using the presented discrete model, we examine the onset of cohesive instability, the influence of intrinsic network defects, the evolution of the cavity population, and the development of strand rupture. The analysis begins with hydrostatic expansion and then considers failure under confined poker-chip loading.

\subsection{Cohesion transforms homogeneous swelling into localized cavitation}
\label{subsec:cohesion_onset}

We first subject the two-dimensional network to quasi-static equi-biaxial expansion and report the normalized hydrostatic tension $\sigma_h\nu/k_BT$ versus in-plane area ratio $J=A/A_0$. Without the van der Waals interaction, the stress increases smoothly  over $1\leq J\leq4$ [Fig.~\ref{fig:cohesion_onset}(a)]. The configurations at $J=1$, 2, and 4 show broadly distributed mesh expansion and no localized cavity appears over this range [Fig.~\ref{fig:cohesion_onset}(b)]. The noncohesive network therefore accommodates expansion through distributed swelling and in this case its bulk modulus scales with the shear modulus, $K\sim G\sim c_0k_BT$, where $c_0$ is the reference strand density.

\begin{figure}[t!]
    \centering
    \includegraphics[width=1\linewidth, trim={1.7cm 0 0 0}]{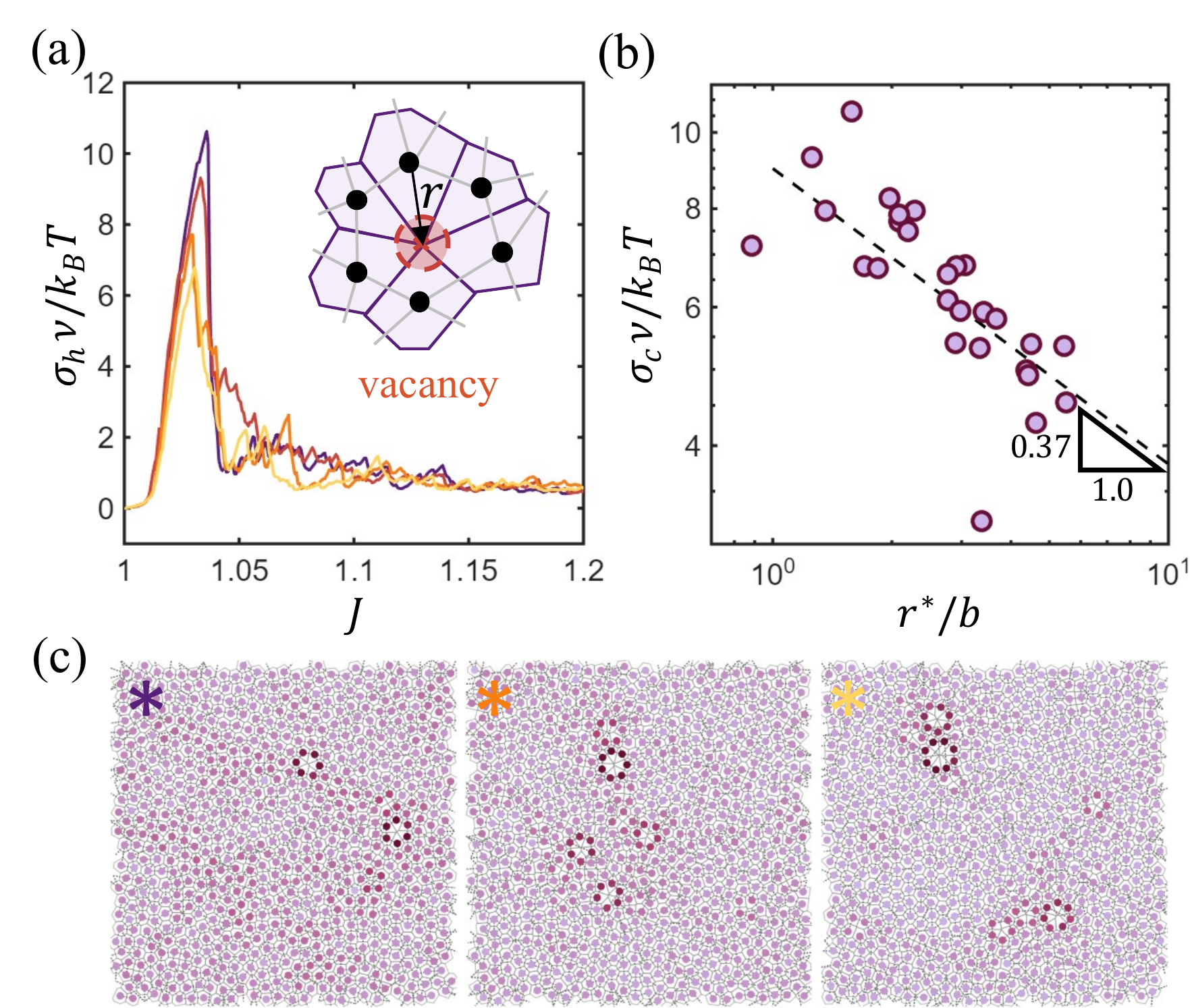}
    \caption{Intrinsic network defects select cavitation onset. (a) Hydrostatic responses of network realizations with different initial defect populations. The inset illustrates the Voronoi construction. (b) Critical hydrostatic tension versus the excess size of the largest initial defect, $r^*=r_{\max}-\langle r\rangle$. The dashed line is a power-law fit, $\sigma_c\nu/k_BT\sim(r^*/b)^{-0.37}$. (c) Representative undeformed networks colored by the local vacancy radius.}
    \label{fig:defect_onset}
\end{figure}
Introducing cohesion produces a much larger initial bulk modulus and a hydrostatic-tension maximum, $\sigma_c=\sigma_h(J^*)$, followed by a stress drop. Increasing $\epsilon_a/k_BT$ raises both the initial stiffness and the critical tension while shifting the instability point toward smaller dilation. These trends are consistent with the cohesive instability of the crosslinked van der Waals description~\cite{cohesive_instability_sam}. Once the dense phase can no longer stabilize a local density fluctuation, expansion concentrates in defects and vacancies.

Figure~\ref{fig:cohesion_onset}(b) shows localized cavity growth after the stress peak. At $\epsilon_a/k_BT=8$, numerous network-scale openings appear and enlarge across the specimen. At $\epsilon_a/k_BT=30$, a small precursor forms near the peak and expands into one dominant cavity. Highly stretched strands surround its boundary, while most of the remote network remains weakly deformed. Rupture of these boundary strands permits further cavity expansion. Cohesion thus affects both the hydrostatic stress required for cavity formation and the number and spatial distribution of cavities after onset.

\begin{figure}[t!]
    \centering
    \includegraphics[width=1.0\linewidth, trim={1.5cm 0 0.5cm 0}]{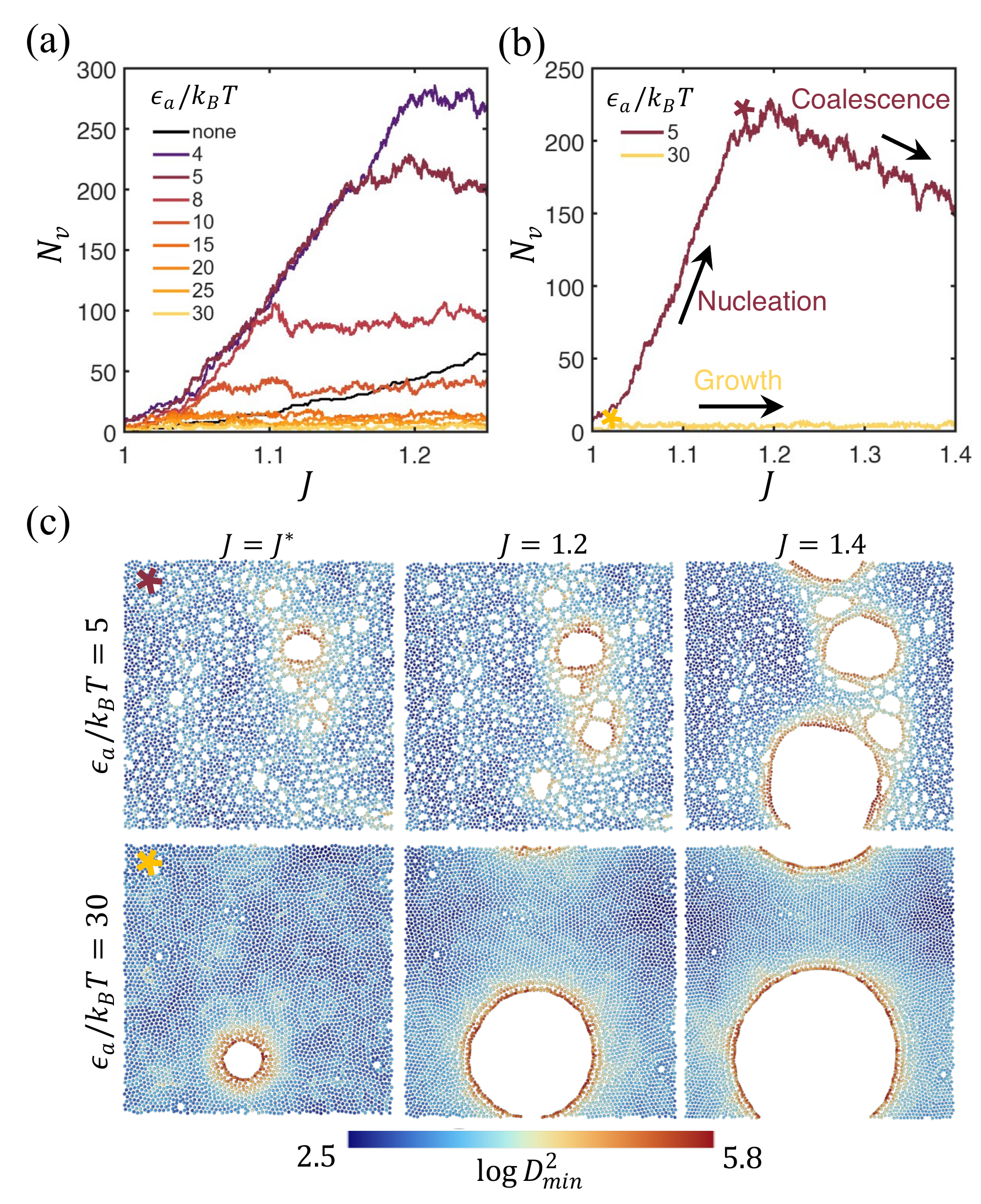}
    \caption{Cohesion controls the number and subsequent interaction of cavities. (a) Number of voids versus volumetric dilation across the complete cohesion series. (b) Number of void regions $N_v$ versus $J$ for weak and strong cohesion of $\epsilon_a/k_BT = 5$ and $\epsilon_a/k_BT = 30$. (c) Configurations at $J=J^*$, 1.2, and 1.4 for $\epsilon_a/k_BT=5$ and 30. Nodes are colored by $\log D^2_{\min}$.}
    \label{fig:cavity_population}
\end{figure}
\begin{figure*}[t!]
    \centering
    \includegraphics[width=1\linewidth]{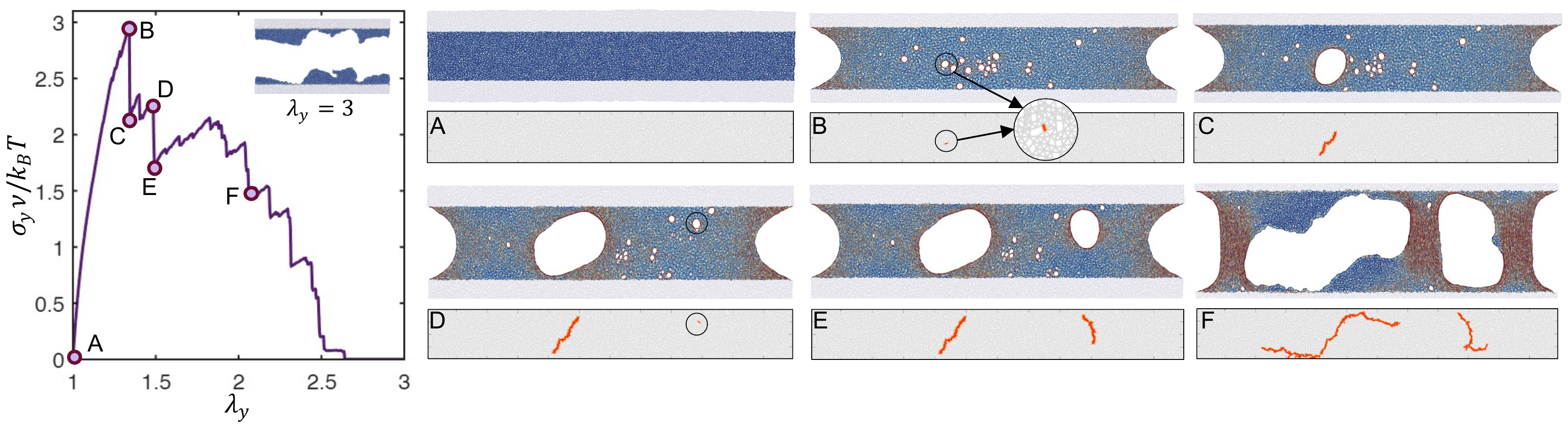}
    \caption{Transition from cavity growth to rupture-mediated failure in a tack-like layer. Normalized tensile stress versus vertical stretch. States A--F denote different points on the curve. The inset illustrates the completely ruptured layer at $\lambda_y=3$. On the right, network configurations at states A--F and corresponding maps of accumulated broken strands. The magnified view at state B identifies the first broken strand adjacent to a small precursor.}
    \label{fig:tack_evolution}
\end{figure*}

\subsection{Intrinsic network defects select the cavitation threshold}
\label{subsec:defect_onset}

Although cohesion provides resistance to dilation and promotes localization, the initial network disorder influences where a cavity first forms and the far-field tension required to activate it. To isolate this effect, we generate networks with identical topology but different initial concentrations of void precursors by varying the equilibration process. Network equilibration consists of overdamped relaxation of the nodes under a sequence of progressively smaller damping coefficients. Void precursor concentration is then increased by decreasing the duration of each damping step. This effectively increases the quench rate, arresting nodes into metastable states and resulting in a higher concentration of pre-cursors. To quantify this initial heterogeneity, we construct a Voronoi tessellation from the nodal positions at $J=1$ [inset of Fig.~\ref{fig:defect_onset}(a)]. A bond-free Voronoi vertex is treated as the center of a vacancy-like region where it lies within an open part of the network through which no strand passes, and its distance to the nearest surrounding node defines the defect radius $r$. This construction identifies unusually large, locally unsupported regions that remain after network equilibration, without prescribing a continuum-scale cavity. Figure~\ref{fig:defect_onset}(c) shows representative initial networks colored by $r$, making the extreme defects visible against the background distribution.

Holding the cohesive strength fixed at $\epsilon_a/k_BT=30$ isolates the effect of this initial structure. The network realizations in Fig.~\ref{fig:defect_onset}(a) exhibit substantially different peak stresses even though they share the same strand and cohesive parameters. We characterize the extreme defect in each realization by $r^*=r_{\max}-\langle r\rangle$, where $r_{\max}$ is the largest vacancy radius and $\langle r\rangle$ is the mean radius in that network. The subtraction distinguishes an anomalously large weak region from a uniformly more open mesh. As shown in Fig.~\ref{fig:defect_onset}(b), the critical hydrostatic tension decreases systematically with $r^*$, and the data follow the empirical relation $\sigma_c\nu/k_BT\sim(r^*/b)^{-0.37}$ over the simulated range. Networks with larger bond-free regions reach cohesive instability at lower far-field hydrostatic tension. This dependence is consistent with the experimentally observed sensitivity of cavitation onset to unresolved defect populations~\cite{DOLLHOFER20046111,Chiche2005}. The responses approach a similar post-peak range, indicating a stronger influence of initial disorder on onset than on subsequent growth.

\subsection{Cohesion governs cavity multiplicity and coalescence}
\label{subsec:cavity_population}

We quantify cavity multiplicity by the number of connected void regions, $N_v$. Figure~\ref{fig:cavity_population}(a) plots $N_v$ versus dilation ratio $J$ across the cohesion series. The noncohesive count represents geometrically enlarged mesh openings and not necessarily localized cavities. Among cohesive networks, increasing cohesion level $\epsilon_a/k_BT$ suppresses cavity multiplicity and localizes the imposed expansion into fewer large cavities. The weak- and strong-cohesion responses are compared directly in Fig.~\ref{fig:cavity_population}(b). At $\epsilon_a/k_BT=5$, the cavity count rises above 200 and subsequently decreases as neighboring cavities coalesce. Its maximum marks a crossover toward coalescence-dominated evolution. However, at $\epsilon_a/k_BT=30$, the count remains small while one dominant cavity expands and no coalescence region is observed.

The maps of non-affine displacement $D^2_{\min}$ in Fig.~\ref{fig:cavity_population}(c) show the spatial distribution of non-affine rearrangement. At weak cohesion, rearrangement occurs around many cavities. Their deformation fields overlap, the intervening ligaments fail, and the cavities merge, reducing $N_v$. At strong cohesion, elevated $D^2_{\min}$ is concentrated around one expanding cavity, while the remote network remains near affine. These observations show that cohesive strength affects both the cavitation threshold and the distribution of subsequent cavity growth, while intrinsic network defects influence the location of the first unstable region.

\subsection{Transition from cavity growth to crack propagation and fracture}
\label{subsec:tack_rupture}

We next examine cavity growth and strand rupture in a bonded, tack-like layer subjected to separation in the $y$ direction. Figure~\ref{fig:tack_evolution} shows the normalized tensile stress $\sigma_y\nu/k_BT$ versus vertical stretch $\lambda_y$. Rising from the equilibrium state A, the stress reaches its global maximum at state B and drops sharply as a cavity opens at state C. Further loading raises the stress again toward state D, followed by another drop as a second cavity opens at state E. These repeated loading and stress-release events involve stretching of the surviving chains around the cavities, local strand rupture, and advancement of cavities and crack-like regions.

Snapshots in figure~\ref{fig:tack_evolution} show the network configurations and accumulated strand rupture at these states. At state B, the magnified region identifies the first broken strand adjacent to a small precursor, while the rapid cavity expansion between states B and C is accompanied by rupture of surrounding strands and the development of a crack-like path. A second precursor becomes active between states D and E, producing another cavity and rupture path. By state F, the cavities have enlarged and the load-bearing ligaments have narrowed, with rupture extending across substantial portions of the layer. These observations connect cavity expansion to progressive loss of network connectivity, consistent with experimental evidence of molecular-scale chain scission around cavities and their evolution into crack-like regions~\cite{Morelle2021,Barney2022Linking}.

\subsection{Decreasing layer thickness distributes damage and extends deformation}
\label{subsec:height_effect}

\begin{figure*}[b!]
    \centering
    \includegraphics[width=1\textwidth, trim={1.0cm 0 0 0}]{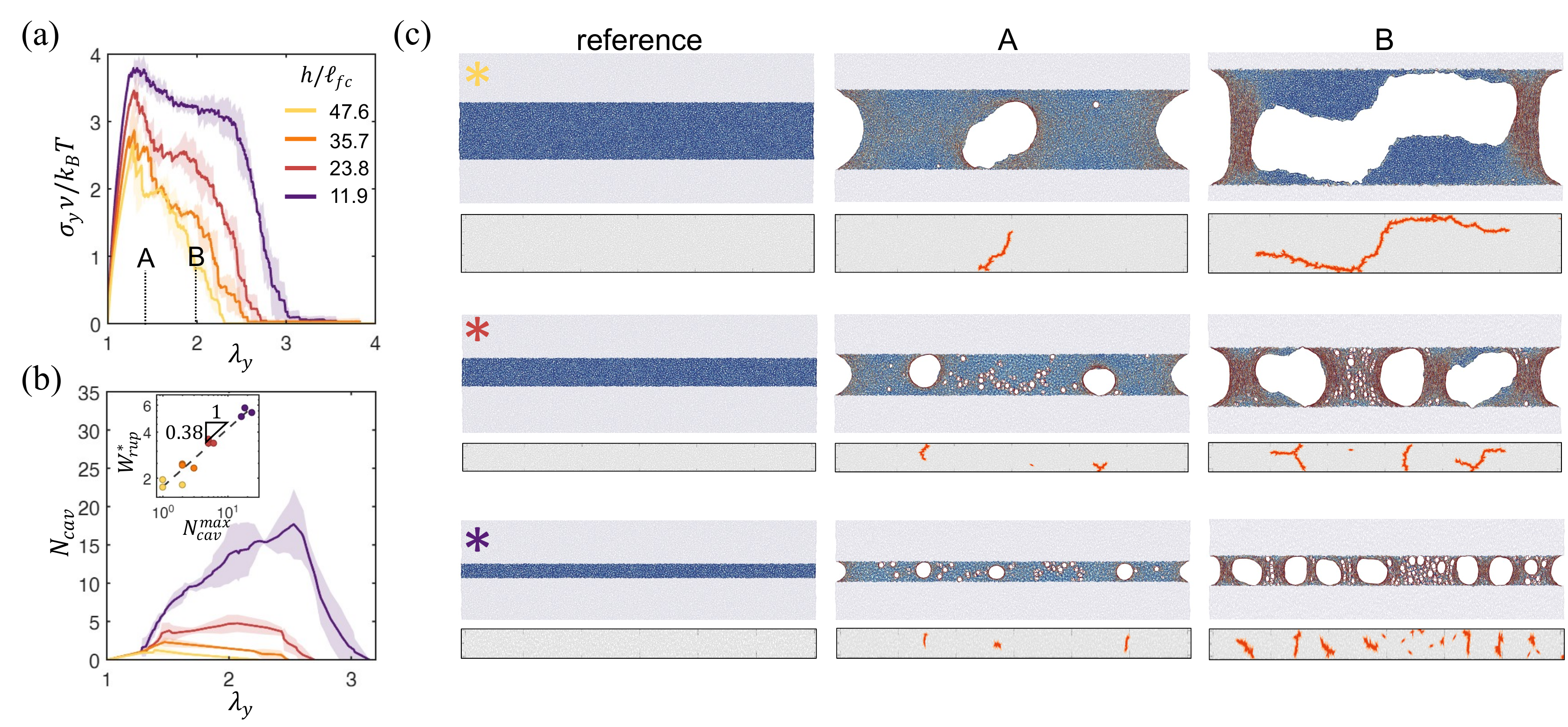}
    \caption{Effect of normalized layer height on tack failure. (a) Normalized tensile stress versus vertical stretch for $h/\ell_{fc}=47.6$, 35.7, 23.8, and 11.9. The shaded bands denote the standard deviation over realizations. (b) Corresponding number of cavities $N_{\mathrm{cav}}$. The inset plots the normalized critical work density of rupture  $W^\star_{\mathrm{rup}}=W_{\mathrm{rup}}\nu/k_BT$ against the maximum cavity count $N_{\mathrm{cav}}^{\max}$ for individual network realizations. The fitted relation is $W^\star_{\mathrm{rup}}\propto(N_{\mathrm{cav}}^{\max})^{0.38}$. (c) Representative configurations at $\lambda_y=1.0$, 1.4, and 2.0 for the heights marked by the yellow, red, and purple stars. Each network image is followed by its accumulated broken-strand map. Thick layers localize into one dominant failure zone, whereas thinner layers develop multiple cavities and spatially distributed rupture. All tack test simulations are performed at $\epsilon_a = 20k_BT$.}
    \label{fig:height_effect}
\end{figure*}

To complete the study, we next examine the effect of confinement by varying the initial layer height $h$ at fixed network parameters. These simulations are performed by clamping two strips of material on top and bottom with thickness $h_{\mathrm{clamp}}$ and applying the tensile stretch by displacing the clamps. Each simulation height is repeated over three network realizations. Experiments on confined elastomers have shown that layer thickness affects the number and spatial distribution of cavities~\cite{Gent1959,LeMenn2022}. We normalize $h$ by the network fractocohesive length $\ell_{fc}=\Gamma/W_c$, where $\Gamma$ is the fracture toughness and $W_c$ is the work per unit volume required to rupture an unnotched network~\cite{Long2020TheFO}. The material length $\ell_{fc}$ remains fixed across the height series. Figure~\ref{fig:height_effect}(a) shows that thinner layers reach higher peak normal stresses and remain load-bearing to larger stretches before complete separation, enclosing a larger area under the stress--stretch curve. We quantify this response through the critical work density of rupture
\begin{equation}
\label{eq:work_rupture}
W_{\mathrm{rup}}
=
\int_{1}^{\lambda_{y,\mathrm{sep}}}
\sigma_y(\lambda_y)\,\mathrm d\lambda_y,
\end{equation}
where $\sigma_y$ is the nominal normal stress and $\lambda_{y,\mathrm{sep}}$ is the stretch at complete separation.

The cavity population increases as the layer becomes thinner [Fig.~\ref{fig:height_effect}(b)]. The thickest layer develops one dominant cavity, while the peak count reaches several cavities at intermediate heights and approximately 18 in the thinnest layer. The subsequent decrease in $N_{\mathrm{cav}}$ results from cavity coalescence and failure of the intervening ligaments. The configurations in Fig.~\ref{fig:height_effect}(c) show how this change in cavity population affects the distribution of damage. At $\lambda_y=1.4$, deformation in the thick layer is concentrated around one principal cavity and rupture path. The thinner layers contain several separated cavities. At $\lambda_y=2.0$, the thinnest layer continues to carry load through narrow ligaments between cavities, with strand rupture distributed across multiple paths. Repeated local failure and reloading of the remaining ligaments accommodate further deformation before complete separation.

This thickness dependence is consistent with a competition between propagation of existing rupture regions and nucleation of additional cavities. For a sufficiently long crack in an ideal pure-shear specimen, the energy release rate scales as $\mathcal G\simeq W_\infty h$, where $W_\infty$ is the recoverable strain-energy density ahead of the crack~\cite{Zhu2024PureShear}. Using $\mathcal G_c=\Gamma=W_c\ell_{fc}$ gives
\begin{equation}
\label{eq:height_energy_release}
\frac{\mathcal G}{\mathcal G_c}
\simeq
\frac{W_\infty}{W_c}\frac{h}{\ell_{fc}}.
\end{equation}
At comparable $W_\infty$, thicker layers therefore provide a larger driving force for propagation, favoring the development of a dominant rupture path. Thinner layers may require additional loading to reach $\mathcal G_c$, during which hydrostatic tension can activate other cavity precursors before an existing rupture path spans the layer. This long-crack estimate provides an energetic interpretation of the observed thickness trend, while the nucleation of microscopic precursors remains governed by the local network response.

The inset of Fig.~\ref{fig:height_effect}(b) connects this distributed failure process to the critical work density of rupture. Across individual network realizations, $W^\star_{\mathrm{rup}}$ increases with the maximum cavity count according to the empirical relation $W^\star_{\mathrm{rup}}\propto(N_{\mathrm{cav}}^{\max})^{0.38}$ over the sampled range. Greater work thus accompanies a larger cavity population and sustained loading of the surviving ligaments. This accumulated external work is distinct from the recoverable density $W_\infty$ and from the fixed unnotched-network rupture density $W_c$ used to define $\ell_{fc}$.

The transition from one dominant cavity in thick layers to multiple cavities that persist over a wider loading interval agrees qualitatively with confined-layer experiments~\cite{Gent1959,Chiche2005,LeMenn2022}, although the onset stresses require a more careful comparison. Le Menn reported no significant thickness dependence of the local hydrostatic stress at first optical cavitation within the experimental scatter~\cite{LeMenn2022}. The increase in macroscopic peak tensile stress observed in our simulations concerns a different lengthscale and need not correspond to the same stage of cavity development. The agreement therefore concerns primarily the more numerous, persistent cavities and the larger stretch to separation in thin layers.

\section{Conclusions}
\label{sec:conclusions}
The discrete simulations connect cavity formation and growth to local cohesion, network disorder, and strand rupture. Under hydrostatic expansion, noncohesive networks accommodate dilation through nearly affine swelling, while cohesive interactions produce a tension maximum followed by localized cavity growth. Increasing cohesive strength raises the critical hydrostatic tension and concentrates expansion into fewer cavities, whereas weaker cohesion permits distributed cavity formation followed by coalescence. At fixed cohesion, the critical tension decreases with the excess size of the largest initial vacancy-like region, linking the onset of cavitation to intrinsic network heterogeneity.

Under confined tack-like loading, cavity opening and extension of rupture paths produce stress drops as highly stretched strands break, followed by reloading of the surviving ligaments. Decreasing the layer height increases the cavity population and distributes strand rupture across multiple regions, allowing thinner layers to remain load-bearing to larger stretches and attain a higher critical work density of rupture. These observations are consistent with an energetic competition in which reduced layer height limits the driving force for propagation of developed crack-like regions, allowing additional cavities to form during continued loading. The transition toward distributed damage agrees qualitatively with confined-layer experiments, although the macroscopic peak normal stress and the local hydrostatic stress at first cavitation remain distinct measures.

Cavitation resistance therefore depends on both the nucleation threshold and subsequent damage evolution, with cohesive strength and layer geometry affecting these stages in different ways. The present two-dimensional, quasi-static model resolves their interaction while omitting three-dimensional defect geometry, viscoelastic rate dependence, and a chemically calibrated strand-scission law. Incorporating these effects would support quantitative comparisons with specific elastomers and adhesives.

\section*{Conflicts of interest}
There are no conflicts to declare.
\section*{Acknowledgements}
F. J. V. gratefully acknowledges the support of the 3M Company under sponsored research contract L-069903-L6-225 and National Science Foundation under Award No. 2023179. This content is solely the responsibility of the authors and does not necessarily represent the official views of the University of Colorado Boulder, 3M Company and National Science Foundation.

\bibliographystyle{elsarticle-num}
\bibliography{Cavitation}

\end{document}


\maketitle
\vspace{-1.2em}

This Supporting Information documents the numerical implementation and loading protocols used in the discrete-network calculations, verifies the local-density estimator and the homogeneous response against the crosslinked van der Waals theory, and provides additional three-dimensional, non-affine, and cavity-shape analyses that support the mechanisms discussed in the main text.

\section{Numerical implementation and analysis}
\label{sec:numerics}

\subsection{Network construction and reduced parameters}

Crosslinking junctions are represented by nodes at positions $\bm r_i$, and a polymer strand joining nodes $i$ and $j$ has connector $\bm r_{ij}=\bm r_j-\bm r_i$ and normalized extension $\lambda_{ij}=r_{ij}/(n_{ij}b)$. The elastic and local-density--dependent cohesive forces are those defined in the main text. The principal calculations are two-dimensional, with $b$ as the length scale and $k_BT$ as the energy scale; the volume per Kuhn segment is therefore interpreted as an effective area, $\nu\sim b^2$. Accordingly, $V$ and $V_c$ represent areas in the two-dimensional
implementation, and $n_q$ has units of inverse area. The auxiliary calculation in Sec.~\ref{sec:three_dimensional} uses the corresponding three-dimensional representation.

For the continuum-validation ensemble, 20 independent networks were generated with 1170 nodes initially distributed in a square domain of side $400b$. Each realization contained 2290 strands with $n_{ij}=20$ Kuhn segments per strand. These choices give the strand-to-segment ratio $\chi=0.05$. Unless a parameter sweep is shown explicitly, the continuum-validation response uses $\epsa/\kBT=20$. The networks were relaxed in the absence of external loading before the reference state $J=1$ was assigned. A strand was removed irreversibly when $\lambda_{ij}\geq\lambda_{\mathrm{rup}}=0.95$, as in the main text.

The conversion from node count to Kuhn-segment count uses the uniform assignment $\kappa=N_K^{\mathrm{tot}}/N_{\mathrm{node}}$ defined in the main text. For the validation ensemble, $N_K^{\mathrm{tot}}=2290\times20=45800$, giving $\kappa\simeq39.15$. This assignment remains fixed after strand rupture.

\begin{table}[t]
\centering
\caption{Baseline parameters of the two-dimensional continuum-validation ensemble at $\chi=0.05$. The cohesive strength is varied in the cohesion sweeps, while the additional values of $\chi$ correspond to different mean Kuhn-segment counts per strand.}
\label{tab:validation_parameters}
\begin{tabular}{@{}ll@{}}
\toprule
Quantity & Value \\
\midrule
Independent realizations & 20 \\
Nodes per realization & 1170 \\
Initial square side length & $400b$ \\
Strands per realization & 2290 \\
Kuhn segments per strand & $n_{ij}=20$ \\
Effective Kuhn-segment area & $\nu=\pi b^2/4$ \\
Baseline strand-to-segment ratio & $\chi=0.05$ \\
Baseline cohesive strength & $\epsa/\kBT=20$ \\
Rupture stretch & $\lambda_{\mathrm{rup}}=0.95$ \\
\bottomrule
\end{tabular}
\end{table}

\subsection{Damped inertial dynamics and time integration}

The nodal trajectories obey damped Newtonian dynamics,
\begin{equation}
m_i\ddot{\bm r}_i=\bm f_i^{\mathrm{el}}+\bm f_i^{\mathrm{vw}}-\zeta\dot{\bm r}_i,
\label{eq:damped_dynamics}
\end{equation}
where $m_i$ is the nodal mass, $\zeta$ is a numerical drag coefficient, $\bm f_i^{\mathrm{el}}=\sum_{j\in\mathcal B_i}\bm f_{ij}^{\mathrm{el}}$, and $\bm f_i^{\mathrm{vw}}$ is the multibody cohesive force. The drag is used to remove inertial transients and relax the network toward a mechanically balanced configuration after each imposed deformation increment; it is not intended as a constitutive model of polymer viscoelasticity.

Equation~\eqref{eq:damped_dynamics} was integrated in LAMMPS~\cite{Thompson2022LAMMPS}. The bonded-particle-model fix was \texttt{nve/bpm/sphere}, whose translational update uses the velocity-Verlet form,
\begin{subequations}
\begin{align}
\bm v_i^{n+1/2}&=\bm v_i^n+\frac{\Delta t}{2m_i}\bm f_i^n,\\
\bm r_i^{n+1}&=\bm r_i^n+\Delta t\,\bm v_i^{n+1/2},\\
\bm v_i^{n+1}&=\bm v_i^{n+1/2}+\frac{\Delta t}{2m_i}\bm f_i^{n+1},
\end{align}
\label{eq:verlet}
\end{subequations}
with the elastic, cohesive, and damping forces reevaluated after the position update. Although the LAMMPS fix is named \texttt{nve}, the complete calculation is not a microcanonical trajectory: energy is removed by the explicit drag, work is supplied through the imposed boundary motion, and strand deletion is irreversible. The time integration is used only as a robust route to successive relaxed states.

\subsection{Loading protocols and reported stresses}

For hydrostatic calculations, the periodic two-dimensional cell was expanded equi-biaxially in small increments and relaxed after each increment. The deformation is reported by the area ratio $J=A/A_0$, and the hydrostatic tension is $\sigma_h=(\sigma_{xx}+\sigma_{yy})/2$, with tension taken as positive. The three-dimensional check uses isotropic expansion with $J=V/V_0$ and $\sigma_h=\mathrm{tr}(\bm\sigma)/3$. The macroscopic stress is obtained from the cell virial of the internal forces after relaxation.

For tack-like calculations, the upper and lower clamped strips were separated in the $y$ direction while the lateral boundaries remained free. The imposed deformation is reported as $\lambda_y=H/h$, where $h$ and $H$ are the initial and current layer heights, respectively. The nominal normal stress $\sigma_y$ is obtained from the clamp reaction using the initial loaded cross section. Consistent with the main text, the critical work density of rupture is
\begin{equation}
W_{\mathrm{rup}}
=
\int_1^{\lambda_{y,\mathrm{sep}}}
\sigma_y(\lambda_y)\,\mathrm d\lambda_y,
\label{eq:work_rupture}
\end{equation}
where $\lambda_{y,\mathrm{sep}}$ denotes complete separation. Its dimensionless form is $W_{\mathrm{rup}}^\star=W_{\mathrm{rup}}\nu/\kBT$. This quantity measures the accumulated external work along the simulated loading path, including the response to cavity growth and irreversible strand rupture.

The effective bulk modulus used in the continuum comparison is the initial tangent $K=(\partial\sigma_h/\partial J)_{J=1}$. For an ensemble quantity, the solid curve or marker denotes the mean and the shaded region denotes one standard deviation when a band is shown. The critical state $J^*$ is identified with the first hydrostatic-stress maximum associated with the abrupt onset of localization.

\subsection{Non-affinity and void-shape measures}

Local non-affine deformation is quantified with the Falk--Langer measure~\cite{FalkLanger1998}. For node $i$, the best local affine tensor $\bm F_i$ maps the reference neighbor vectors to the current configuration, and
\begin{equation}
D^2_{\min,i}(J)=\min_{\bm F_i}\sum_{j\in\mathcal N_i^{D}}\left\lVert\bm r_{ij}(J)-\bm F_i\bm r_{ij}(1)\right\rVert^2,
\label{eq:dmin}
\end{equation}
where $\mathcal N_i^{D}$ is the analysis neighborhood in the reference configuration. The network average is $\langle D^2_{\min}\rangle=N_{\mathrm{node}}^{-1}\sum_iD^2_{\min,i}$. A locally affine deformation gives $D^2_{\min,i}=0$, whereas a large value identifies rearrangement that cannot be represented by one local deformation gradient. Lengths are expressed in units of $b$, so the plotted squared displacements are in units of $b^2$. Logarithmic color scales are applied to the dimensionless quantity $D^2_{\min}/b^2$.

Void number and shape were evaluated from the connected void regions using the same segmentation and connectivity convention at every deformation state. For each connected region, let $\mu_1\geq\mu_2$ be the eigenvalues of its second central-moment tensor. Its eccentricity is
\begin{equation}
e=\sqrt{1-\frac{\mu_2}{\mu_1}}.
\label{eq:eccentricity}
\end{equation}
Equal principal second moments give $e=0$, while $e\rightarrow1$ indicates an increasingly elongated region. The reported $\langle e\rangle$ is the arithmetic mean over the connected regions present in that frame and is defined only when at least one region is detected.

\section{Validation of the local-density estimator}
\label{sec:density_validation}

The cohesive energy depends on a smoothed estimate of the number of nodes, and hence of the number of Kuhn segments, within a cutoff neighborhood. We therefore tested the estimator independently of the network mechanics. With the normalized compact kernel $\widehat\phi(r)$ defined in the main text, the weighted local node count and corresponding $d$-dimensional number density are
\begin{equation}
\widetilde N_i^{\mathrm{node}}=\sum_{j\in\mathcal N_i}\widehat\phi(r_{ij}),\qquad \rho_i^{\mathrm{local}}=\frac{\widetilde N_i^{\mathrm{node}}}{V_d(r_c)},
\label{eq:local_density}
\end{equation}
where $V_1(r_c)=2r_c$ and $V_2(r_c)=\pi r_c^2$. Kernel normalization makes the estimator recover the imposed density for a spatially uniform distribution in the continuum limit.

\subsection{One-dimensional test}

\begin{figure}[H]
\centering
\includegraphics[width=\linewidth]{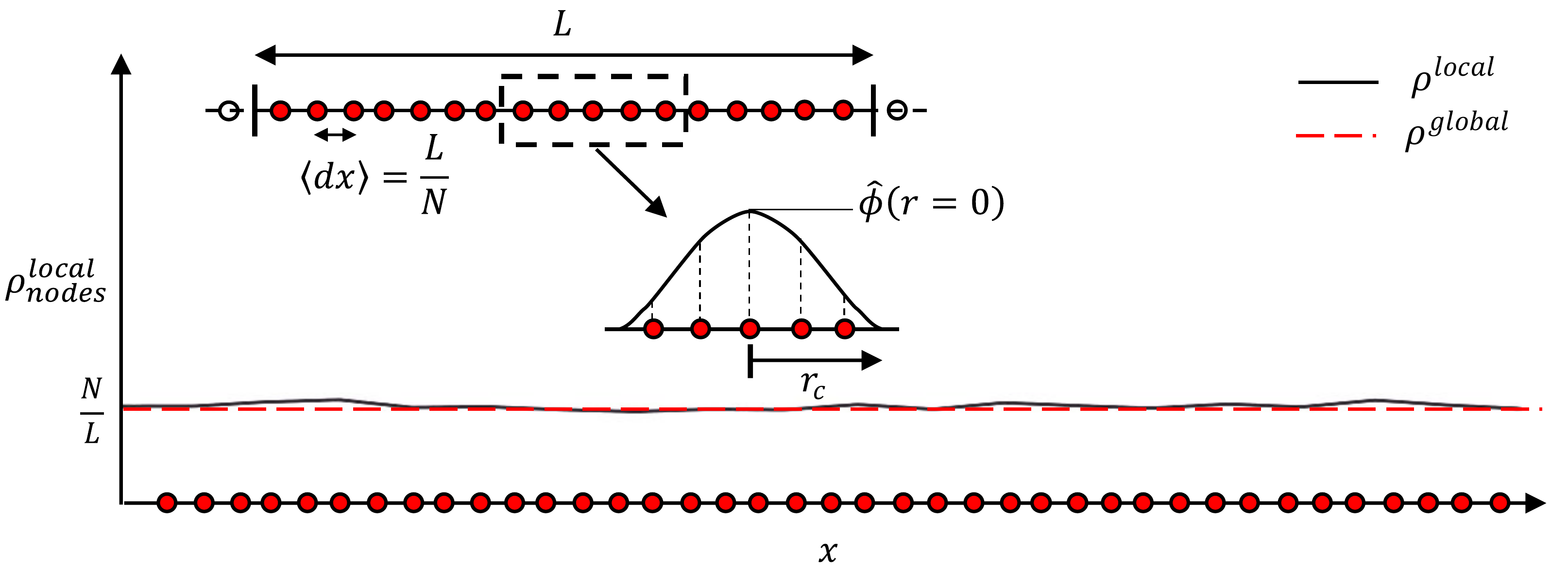}
\caption{One-dimensional validation of the local-density estimator. Uniformly spaced nodes occupy a line of length $L$, and the inset illustrates the compact weighting kernel within $r_c$. The local estimate $\rho_i^{\mathrm{local}}=\widetilde N_i^{\mathrm{node}}/(2r_c)$ (black) closely follows the exact global density $N/L$ (red dashed).}
\label{fig:density_1d}
\end{figure}

For $N$ uniformly spaced nodes in a line of length $L$, the exact density is $\rho^{\mathrm{global}}=N/L$. Figure~\ref{fig:density_1d} compares this value with Eq.~\eqref{eq:local_density}. The local estimate remains close to $N/L$ along the entire line; the small fluctuations reflect the small perturbation we add to nodal positions.

\subsection{Two-dimensional test}

\begin{figure}[H]
\centering
\includegraphics[width=\linewidth]{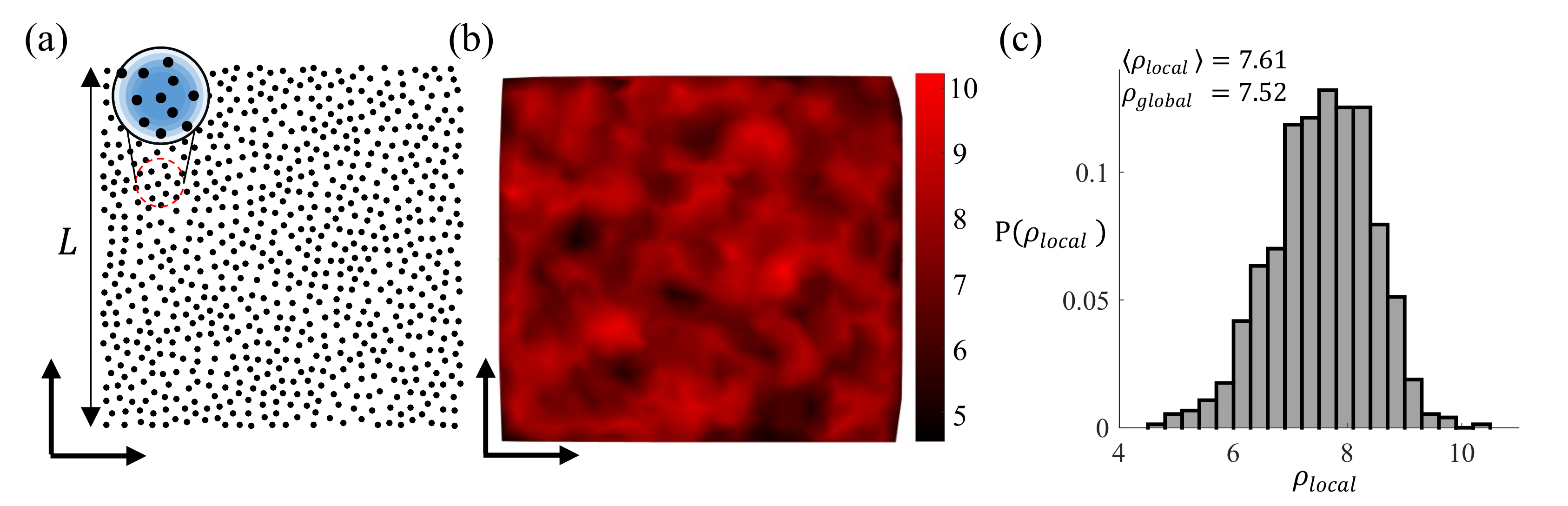}
\caption{Two-dimensional validation of the local-density estimator. (a) Disordered node distribution and a representative cutoff neighborhood. (b) Spatial interpolation of $\rho_i^{\mathrm{local}}=\widetilde N_i^{\mathrm{node}}/(\pi r_c^2)$. (c) Distribution of local estimates. The measured mean, 7.61, differs from the exact global density, 7.52, by $1.2\%$.}
\label{fig:density_2d}
\end{figure}

For $N$ disordered nodes in a square of side $L$, the exact global density is $\rho^{\mathrm{global}}=N/L^2$. Figure~\ref{fig:density_2d}(a) shows the test distribution and one representative cutoff neighborhood, while Fig.~\ref{fig:density_2d}(b) shows the spatially interpolated local estimate. The histogram in Fig.~\ref{fig:density_2d}(c) is centered at $\langle\rho^{\mathrm{local}}\rangle=7.61$, compared with $\rho^{\mathrm{global}}=7.52$. The relative difference is $1.2\%$. The measured mean is therefore close to the imposed global density for this test distribution, while the local estimates vary with the spatial arrangement of nodes.

\section{Validation against the homogeneous continuum theory}
\label{sec:continuum_validation}

\begin{figure}[H]
\centering
\includegraphics[width=\linewidth]{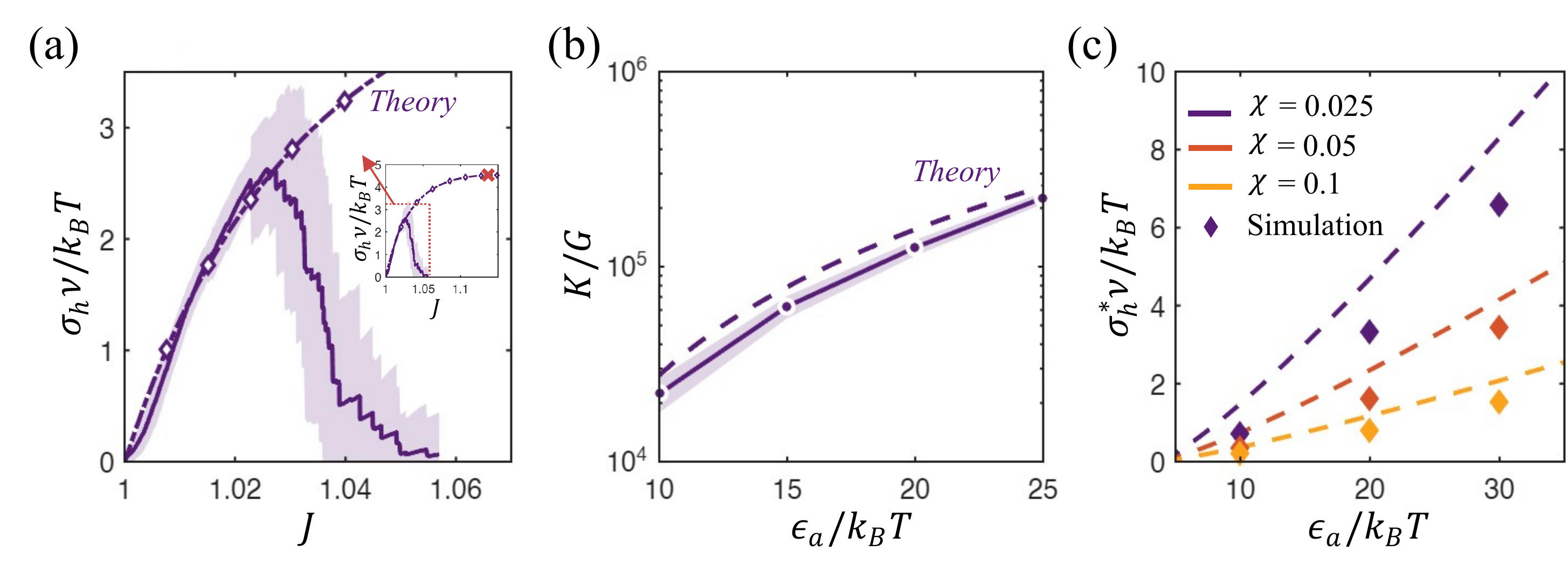}
\caption{Validation of the discrete cohesive model against the homogeneous crosslinked van der Waals theory. (a) Hydrostatic tension versus area ratio for $\chi=0.05$ and $\epsa/\kBT=20$; the main panel magnifies the pre-instability response and the inset shows the full curve. The continuum prediction is shown by the dashed curve with diamonds, and the discrete ensemble by the solid curve and shaded band. (b) Effective bulk modulus versus cohesive strength at fixed $\chi=0.05$. (c) Critical hydrostatic tension versus cohesive strength for $\chi=0.025$, 0.05, and 0.1. Dashed curves are continuum predictions and diamonds are discrete results. Bands in (a,b) show one standard deviation over 20 independent network realizations.}
\label{fig:continuum_validation}
\end{figure}

The local cohesive potential is a discrete realization of the crosslinked van der Waals free energy of Lamont et al.~\cite{cohesive_instability_sam}. Before localization, the continuum comparison is made using $G=c_0\kBT$ and $\chi f_0=\nu c_0$, which give
\begin{equation}
\frac{\sigma_h}{G}=\frac{\sigma_h\nu/\kBT}{\chi f_0},\qquad \frac{K}{G}=\frac{K\nu/\kBT}{\chi f_0}.
\label{eq:continuum_normalization}
\end{equation}
Figure~\ref{fig:continuum_validation}(a,c) reports stresses normalized by $\kBT/\nu$, while panel (b) reports the bulk modulus normalized by $G$. Equation~\eqref{eq:continuum_normalization} relates these two normalizations.

Figure~\ref{fig:continuum_validation}(a) compares the discrete and continuum hydrostatic responses for $\chi=0.05$ and $\epsa/\kBT=20$. The main panel magnifies the initial loading branch and the onset region, while the inset shows the full response. The discrete networks reproduce the steep initial increase in hydrostatic tension, but reach their peak at smaller dilation and lower tension than the homogeneous continuum prediction. Figure~\ref{fig:continuum_validation}(b) also shows a similar dependence of bulk modulus on cohesive strength, with systematically lower values for the discrete networks.

The critical-stress comparison in Fig.~\ref{fig:continuum_validation}(c) shows that both descriptions predict increasing cavitation resistance with $\epsa/\kBT$ and decreasing $\chi$, although the discrete critical stresses remain below the homogeneous predictions. The discrete networks contain variations in local density, strand orientation, and connectivity, and can relax through non-affine rearrangement. These features provide plausible routes to localized instability before the corresponding homogeneous state becomes unstable. The comparison therefore supports qualitative consistency in the cohesion dependence of stiffness and cavitation resistance, while the differences in modulus and onset stress limit quantitative agreement.

\section{Three-dimensional response and comparison with molecular dynamics}
\label{sec:three_dimensional}

\begin{figure}[H]
\centering
\includegraphics[width=\linewidth]{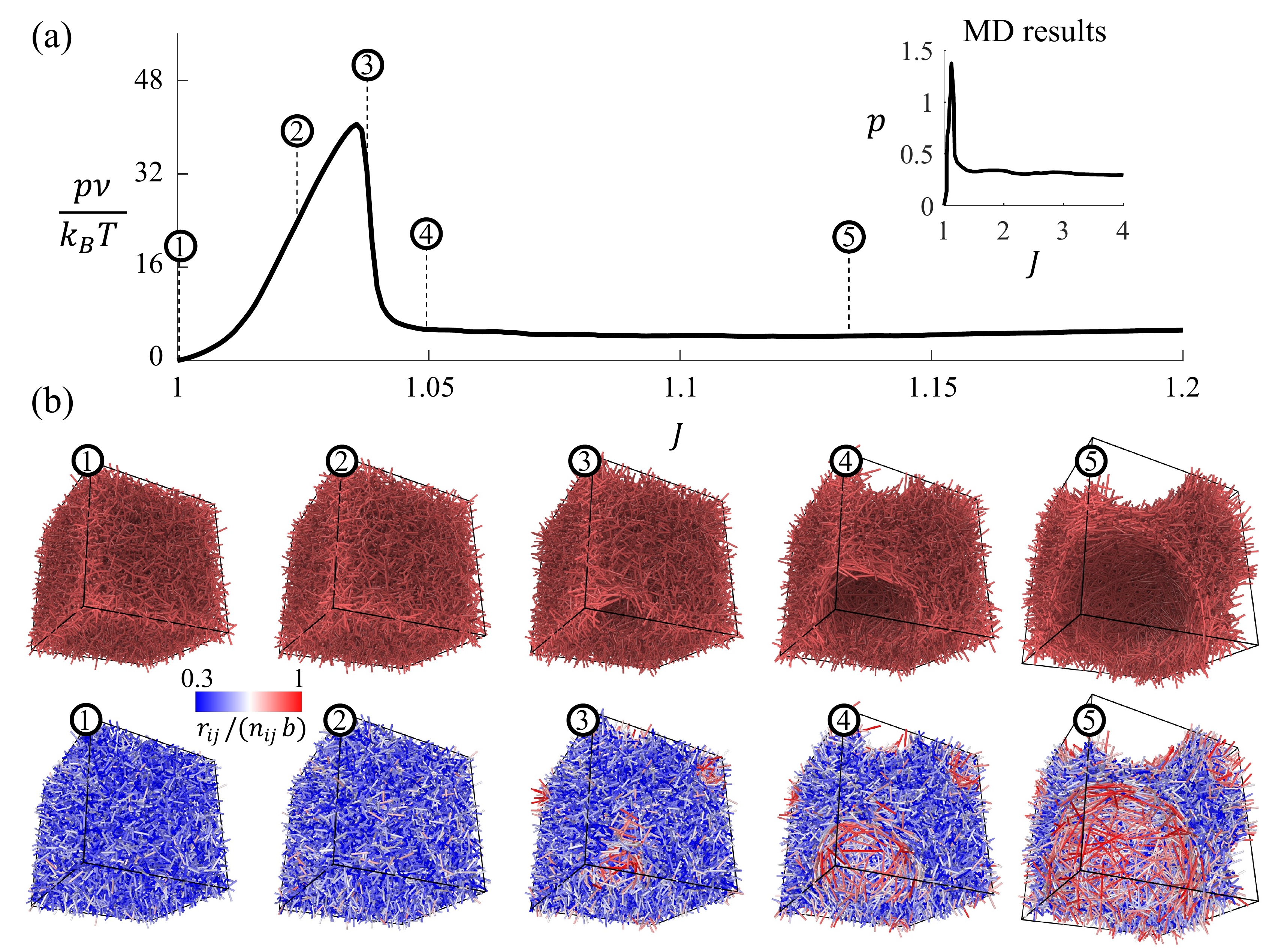}
\caption{Auxiliary three-dimensional calculation and qualitative comparison with coarse-grained molecular dynamics. (a) Discrete hydrostatic response; the inset is a qualitative trace redrawn from the response reported by Ye and Riggleman~\cite{Riggleman2020}. States 1--5 mark the reference state, homogeneous loading, the vicinity of the maximum, post-peak cavity opening, and subsequent cavity growth. (b) Corresponding network configurations. The upper row shows the full geometry in a uniform color; the lower row colors strands by $\lambda_{ij}=r_{ij}/(n_{ij}b)$ from 0.3 (blue) to 1 (red).}
\label{fig:md_comparison}
\end{figure}

The main study uses two-dimensional networks to make ensemble sampling and extensive parameter sweeps tractable. Figure~\ref{fig:md_comparison} provides an auxiliary three-dimensional calculation. The discrete response rises steeply from the reference state, reaches a maximum near $J\simeq1.04$, drops abruptly as a cavity forms, and then approaches a much lower cavity-supporting stress. The inset shows the same qualitative peak--drop--plateau sequence in the coarse-grained molecular-dynamics calculations of Ye and Riggleman~\cite{Riggleman2020}. Because the two models use different coarse-graining, interaction potentials, and stress normalizations, this comparison concerns the sequence of regimes and not pointwise agreement in pressure or stretch.

The configurations in Fig.~\ref{fig:md_comparison}(b) link the macroscopic response to the network morphology. State 1 is the relaxed reference network, state 2 lies on the homogeneous rising branch, and state 3 is close to the pressure maximum, where the first localization becomes apparent. State 4 follows the stress drop and contains an open internal cavity. Continued dilation enlarges that cavity through state 5. The lower row is colored by $\lambda_{ij}=r_{ij}/(n_{ij}b)$ and shows that the most highly extended strands progressively concentrate around the cavity surface, while much of the remote network remains less stretched. The three-dimensional calculation thus shows cavity formation accompanied by localized strand stretching, qualitatively consistent with the behavior observed in the two-dimensional networks.

\section{Cohesion-dependent non-affine localization}
\label{sec:nonaffine}

\begin{figure}[H]
\centering
\includegraphics[width=0.72\linewidth]{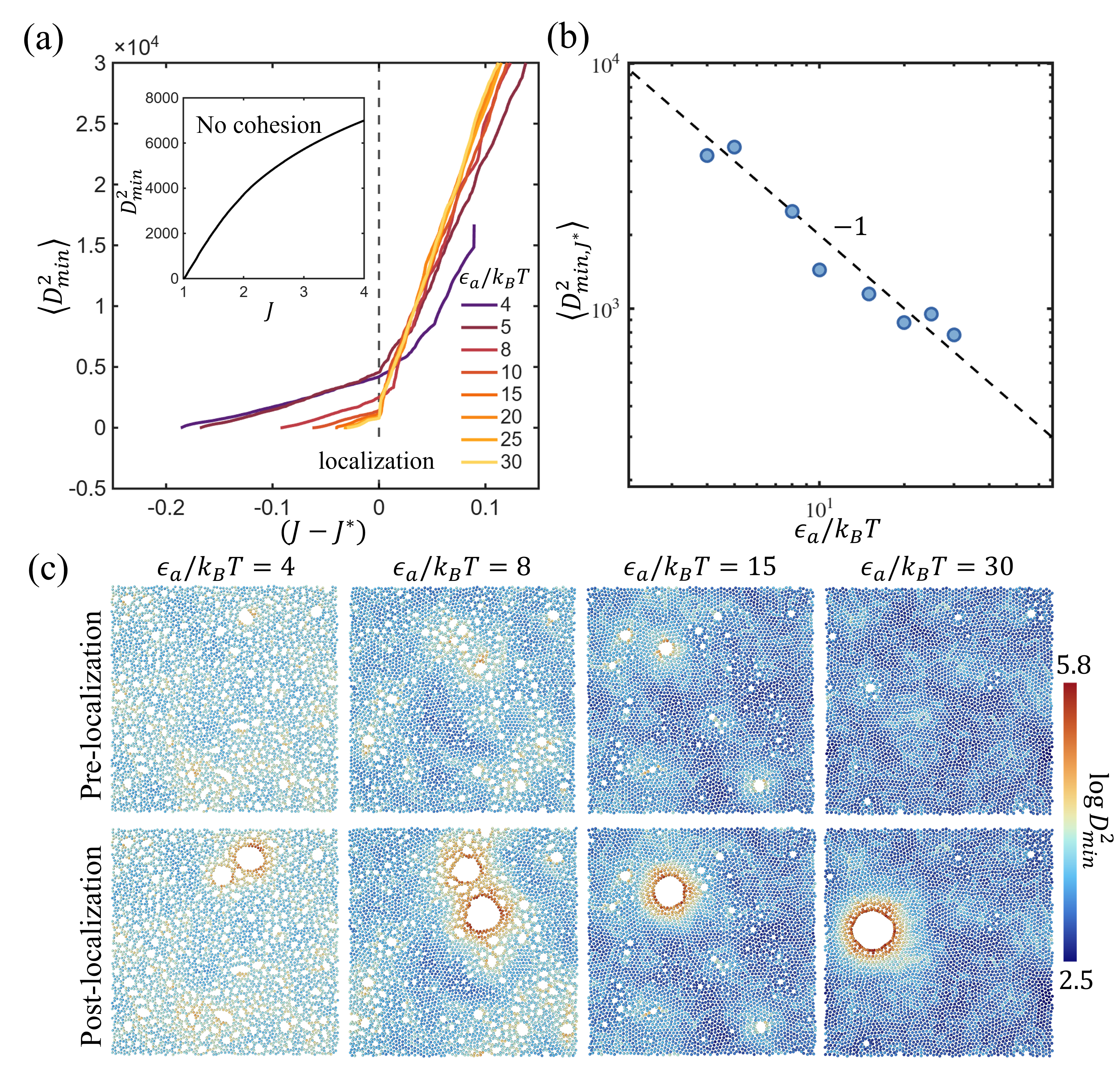}
\caption{Non-affine localization as a function of cohesion. (a) Network-averaged $\langle D^2_{\min}\rangle$ versus $J-J^*$ for $\epsa/\kBT=4$, 5, 8, 10, 15, 20, 25, and 30; the dashed line marks localization, and the inset shows the smooth noncohesive response versus $J$. (b) Log--log plot of the value at localization versus cohesive strength. The dashed guide has slope $-1$. (c) Configurations immediately before and after localization for $\epsa/\kBT=4$, 8, 15, and 30 from left to right, colored by $\log D^2_{\min}$. Weak cohesion activates distributed regions, whereas strong cohesion concentrates non-affinity around one cavity.}
\label{fig:nonaffine}
\end{figure}

Figure~\ref{fig:nonaffine}(a) aligns the cohesive responses at $J-J^*=0$ to compare the evolution of non-affinity around the hydrostatic-stress maximum. For the cohesive cases examined, $\langle D^2_{\min}\rangle$ increases gradually before $J^*$ and more rapidly after it. The noncohesive response is shown against $J$ because no corresponding localization-associated maximum is identified over the simulated interval $1\leq J\leq4$.

The values of $\langle D^2_{\min}\rangle$ at $J^*$ are approximately consistent with an inverse dependence on $\epsa/\kBT$ over the sampled range, as indicated by the slope-$-1$ guide in Fig.~\ref{fig:nonaffine}(b). The configurations in Fig.~\ref{fig:nonaffine}(c) suggest that this trend is associated with a change in the spatial extent of rearrangement. At weak cohesion, elevated non-affinity is distributed around many openings, while stronger cohesion concentrates it around fewer cavities. A smaller fraction of strongly rearranging nodes can reduce the network average even when local values remain large, providing a possible explanation for the observed trend.

\section{Void morphology during nucleation, growth, and coalescence}
\label{sec:void_morphology}

\begin{figure}[H]
\centering
\includegraphics[width=\linewidth]{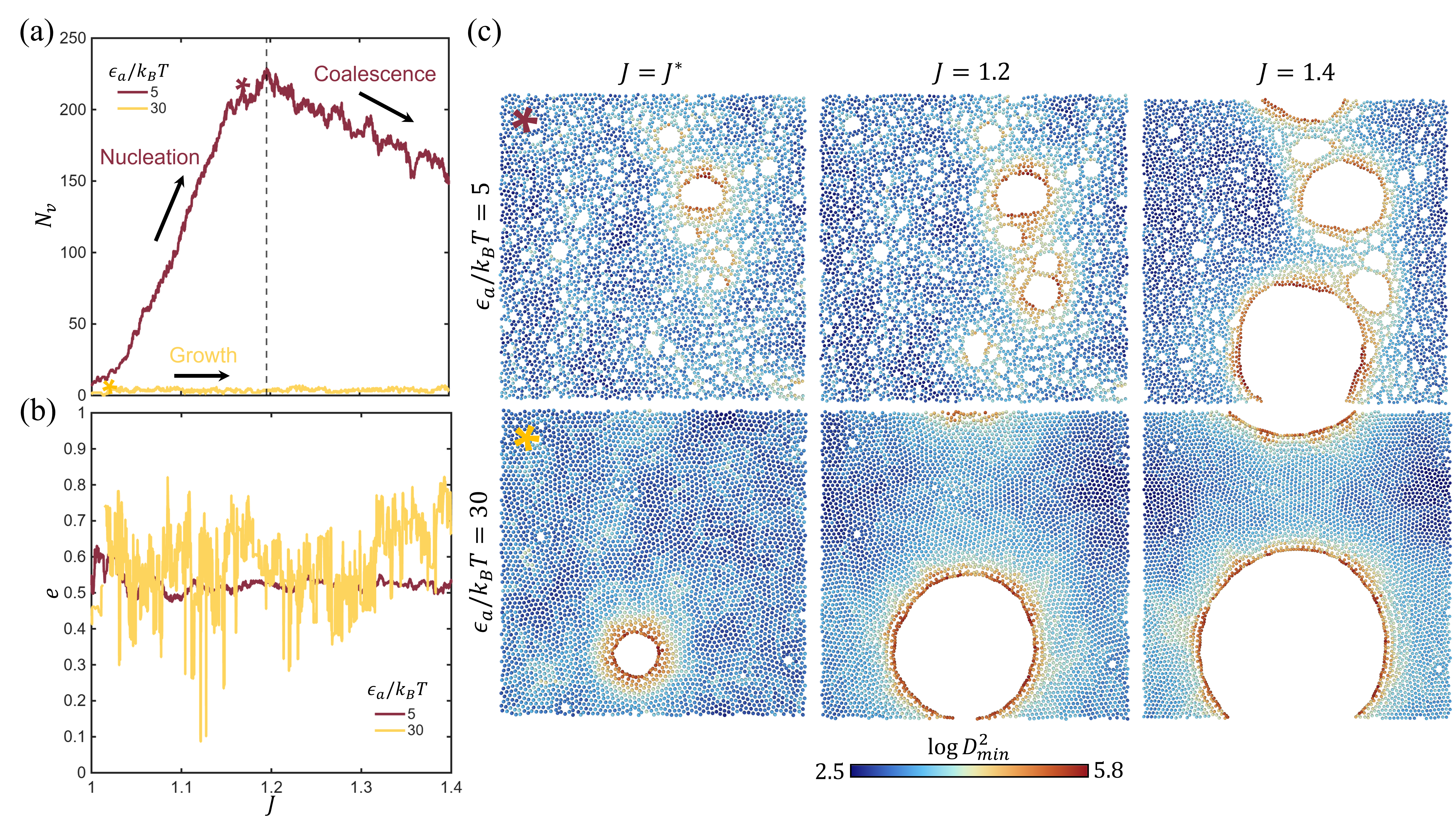}
\caption{Additional cavity-population and shape analysis for weak and strong cohesion. (a) Number of connected void regions versus $J$ for $\epsa/\kBT=5$ (maroon) and 30 (yellow). The weak-cohesion maximum separates the nucleation-dominated and coalescence-dominated regimes; strong cohesion produces growth of essentially one cavity. (b) Mean void eccentricity defined by Eq.~\eqref{eq:eccentricity}. (c) Configurations at $J=J^*$, 1.2, and 1.4 for weak cohesion (upper row) and strong cohesion (lower row), colored by $\log D^2_{\min}$. Stars in (a) identify the corresponding $J^*$ states.}
\label{fig:void_morphology}
\end{figure}

Figure~\ref{fig:void_morphology}(a) compares the two limiting cohesive responses used in the main text. At $\epsa/\kBT=5$, the connected-void count rises rapidly as many network defects open, reaches a maximum near $J\simeq1.2$, and then decreases even though the total dilation continues. The decrease in cavity count, together with the configurations, is consistent with coalescence as neighboring cavities merge and the intervening ligaments disappear. At $\epsa/\kBT=30$, the count remains close to one because dilation is accommodated predominantly by growth of the initially selected cavity.

The mean eccentricity in Fig.~\ref{fig:void_morphology}(b) adds a shape measure to this population count. The many weak-cohesion cavities have a relatively stable mean $\langle e\rangle\simeq0.5$ through nucleation and coalescence. The strong-cohesion trace fluctuates more because the statistic is based on only one or a few regions, but it does not show a sustained progression toward a crack-like, highly elongated shape. The snapshots in Fig.~\ref{fig:void_morphology}(c) are consistent with this distinction: weak cohesion produces several irregular cavities that impinge and merge, whereas strong cohesion produces one nearly rounded cavity whose radius increases while non-affine deformation remains concentrated along its perimeter. These observations indicate that cohesion affects both cavity multiplicity and the spatial distribution of growth in the hydrostatic calculations. The eccentricity describes the geometric anisotropy of the void regions and does not by itself distinguish elastic expansion from rupture-mediated growth.

\clearpage
\bibliographystyle{elsarticle-num}
\bibliography{Cavitation}